\documentclass[12pt]{article}
\usepackage{amsmath}
\usepackage{amsfonts}
\usepackage{amssymb}
\usepackage{graphicx}

\numberwithin{equation}{section}

\begin{document}

\newcommand{\ls}[1]
   {\dimen0=\fontdimen6\the\font \lineskip=#1\dimen0
\advance\lineskip.5\fontdimen5\the\font \advance\lineskip-\dimen0
\lineskiplimit=.9\lineskip \baselineskip=\lineskip
\advance\baselineskip\dimen0 \normallineskip\lineskip
\normallineskiplimit\lineskiplimit \normalbaselineskip\baselineskip
\ignorespaces }

\title{ The Distribution Route from Ancestors to Descendants}
\author{Baruch Fischer and Moshe Zakai\\
Department of Electrical Engineering\\
Technion, Haifa 32000, Israel\\
E-mail: fischer@ee.technion.ac.il ,  zakai@ee.technion.ac.il}

\date{}
\maketitle \thispagestyle{empty} \setcounter{page}{1}

 \ls{1.5}
\begin{abstract}
We study the distribution of descendants of a known personality, or
of anybody else, as it propagates along generations from father or
mother through any of their children. We ask for the ratio of the
descendants to the total population and construct a model for the
route of Distribution from Ancestors to Descendants (DAD).  The
population ratio $r_n$ is found to be given by the recursive
equation $ r_{n+1} \approx (2-r_n) r_n\,,$ that provides the
transition from the $n-$th to the $(n+1)$th generation. $ r_0
=1/N_0$ and $N_0$ is the total relevant population at the first
generation. The number of generations it takes to make half the
population descendants is $\log N_0/\log 2$ and additional $\sim 4$
generations make everyone a descendent (=the full descendant
spreading time). These results are independent of the population
growth factor even if it changes along generations. As a running
example we consider the offspring of King David. Assuming a
population between $N_0 = 10^6$ and $5 \cdot 10^6$ of Israelites at
King David's time ($\sim 1000$ BC), it took 24 to 26 generations
(about 600-650 years, when taking 25 years for a generation) to make
every Israelite a King David descendent. The conclusion is that
practically every Israelite living today (and in fact already at
350-400 BC), and probably also many others beyond them, are
descendants of King David. We note that this work doesn't deal with
any genetical aspect. We also didn't take into account here any
geo-social-demographic factor. Nevertheless, along tens of
generations, about 120 from King David's time till today, the DAD
route is likely to govern the distribution in communities that are
not very isolated.
\end{abstract}

\newpage

\section{Introduction}
The well known Galton-Watson (GW) process \cite{1,2} investigates
the extinction of surnames which propagate from father to son.  We
consider here a model which unlike the GW model depends on both
parents, namely the offspring of a known personality which
propagates through father or mother to their children and we ask for
the ratio of the descendants to the total population. As an example
we will consider the offspring of King David.

King David lived about 3,000 years ago. We assume that a generation
(from birth until marriage and children) is 25 years, and each
married couple has $2g$ children, where $g$ is the growth factor per
generation. Let $N_n$ denote the number of Israelites at the $n-$th
generation and $N_n = N_0 \cdot g^n$ where $n=0,1,2,\cdots,120
(=3000/25)$. Let $\tilde{D}_n$ denote the number of descendants
(male and female) of King David at the $n-$th generation and
$\tilde{C}_n = N_n - \tilde{D}_n$ denote the non-descendants. We
start at the first generation with $\tilde{D}_0=1$, the dynasty
founder. Our problem is to estimate the ratio $r_n=\tilde{D}_n/N_n$
after $n$ generations. This will show, in particular, that
practically all Israelites today are descendants. It is possible
that a family disappears after a few generations (discussed in
Section 6), but we assume throughout the paper that it doesn't
happen. In our example with what we know about King David and his
son Solomon, we do not have to worry about that. However if we are
not sure about that we can say, as we discuss later, that
\emph{either all Israelites are his descendant or none}. Therefore
if there is one descendant then all are descendants.

It will be shown that for a population of $N_0$ it took $\log
N_0/\log 2$ generations to make half the population descendants of
King David. Additional four generations made all of them his
descendants. That is the DAD full spreading time. The transitions
region between low to high spreading ratio is very quick, a few
generations. It is not only King David; the same relation exists
regarding anyone else of his era (or any other early era) whose
family survived in the first few generations (discussed in Section
6), including for example less admirable characters in the Bible
like Nabal... Assuming a population between $N_0 = 10^6$ and $N_0 =
5 \cdot 10^6$ Israelites at King David's time ($\sim 1000$ BC)
\cite{3}, it took 24 to 26 generations (600-650 years, when taking
25 years for a generation) to ensure that every Israelite was his
descendant. That means that every Israelite living at 400 BC, the
beginning of the era of the Second Temple in Jerusalem, was already
a descendant of King David.

An interesting feature of the DAD route is that the descendant
population ratio and the spreading time depend on $N_0$ but not on
the population growth factor $g$ even if $g$ is generation
dependent.

\vspace{0.7cm}

 \section{The rule of passing from $\tilde D_n$ to $\tilde{D}_{n+1}$}

 Two presentations of the same result are given.
 \paragraph{The First Presentation:} Let $N_n$ be the total number of males and females
 at the $n-$th generation of a certain community, and among
 them $\tilde D_n$ (Davidian) descendants of the dynasty
 we follow. The number of males and females are assumed equal. Further assume
 that we have $N_n/2$ cards with all the female names, one name per
 card, and similarly in another box $N_n/2$ cards for all the males.
 Consider now a "match maker" picking up randomly and independently
 one card from each box, and combining them to a single card (marriage) with
 the two names on it. At this stage add to
 each of the two names information whether he (she) is or isn't a
 D-descendant. Throw away all cards where both names are
 non-descendants. The probability that a descendant married a
 descendant is $\tilde D_n/N_n$ and for a descendant
 marrying a non-descendant it is $(N_n - \tilde D_n)/N_n$.

 The conditional expectation $E(\tilde{D}_{n+1} | \tilde{D}_n)$ (the
 number of descendants at the $(n+1)$th generation conditioned on
 $\tilde{D}_n$, the number of descendants at the $n-$th generation)
 follows from the fact that for $g=1$ a married couple who are both
 descendants will have 2 descendants.  In this case the number of
 descendants will not change.  On the other had if a descendant
 married a non-descendant the one descendant in the $n-$th generation
 will generate two descendants for the $(n+1)$th generation.

\noindent Therefore,
\begin{align}
\label{eq21} E(\tilde{D}_{n+1} | \tilde{D}_n) & = g \cdot \left[
\begin{array}{l} \text{the probability that} \\
\text{a descendant married} \\
\text{a descendant}
\end{array}
\right] \cdot \tilde{D}_n + 2g \cdot \left[
\begin{array}{l}
\text{the probability that} \\
\text{a descendant married} \\
\text{a non-descendant}
\end{array}
\right] \cdot \tilde{D}_n\notag\\
&= \left(1\cdot\frac{\tilde {D}_n}{N_n} + 2 \cdot\frac{N_n-
\tilde{D}_n}{N_n} \right)\cdot g \tilde{D}_n= \left(2-
\frac{\tilde{D}_n}{N_n}\right)\cdot g \tilde{D}_n\,.
\end{align}
Note that $\tilde{D}_n$ and $N_n$ include males and females, and we
assume throughout the paper that their numbers are equal. We also
comment that this statistical model can be viewed as a random walk
process, as discussed below in section 3.

We can simplify the analysis by a simple renormalization procedure
that shows right away that the DAD route is independent of the
growth factor $g$. We normalize by setting:
\begin{equation}
 \label{eq22}
 D_n = \frac{\tilde{D}_n}{g^n}\;\;, \;\;\;\; C_n = \frac{\tilde C_n}{g^n}
 = N_0- D_n \;\;,
 \end{equation}
 and then with $N_n = N_0\;g^n$ we obtain from Eq.
\eqref{eq21}
 \begin{equation}
 \label{eq23}
 E (D_{n+1} | D_n) = \left(2-\frac{D_n}{N_0}\right)D_n \,.
 \end{equation}
The growth factor $g$ is eliminated and therefore $N_0$ is the only
relevant quantity. Not only that, but the free of $g$ property holds
for the general case where $g$ depends on $n$ and then $g_n$ is the
growth factor from the generation $n$ to $(n+1)$. To realize that,
we replace in the former procedure: \vspace*{-0.7cm}
\begin{equation}
g\rightarrow g_n\;,\;\;\;\;\label{eq231} g^n \rightarrow (g_0 \cdot
g_1 \cdot g_2...\cdot g_{n-1})=\prod_{i=0}^{n-1} g_i \;.
\end{equation}
\noindent Then $N_n=N_0 \cdot \prod_{i=0}^{n-1}g_i$. Usually $g_n
\ge1$, but at least we need for the first few generations $g_n
> 1/2$ to avoid extinction of $D_n$ (discussed in Section 6), and for later generations $g_n$
to be bounded away from zero. With the normalization:
\vspace*{-0.5cm}
\begin{equation}
\label{eq232} D_n =\tilde D_n/\prod_{i=0}^{n-1} g_i\;,\;\;\;\, C_n
=\tilde C_n /\prod_{i=0}^{n-1} g_i\;.
 \end{equation}
\noindent Eq. \eqref{eq21} with $g\rightarrow g_n\,$ transforms to
Eq. \eqref{eq23} which is free of all $g_n$.

In section 3 we view this statistical process as a simple random
walk model. We will show that $E(D^2_n) \approx E^2(D_n)$, and
therefore the recursive Eq. \eqref{eq23} can be written as:
\begin{equation}
 \label{eq24}
 E(D_{n+1} ) \approx \left(2-\frac{ED_n}{N_0}\right)E D_n \,.
 \end{equation}

 \paragraph{The Second Presentation:} We split the population into
 males and females,
 $N_n^m = N_n^f = \frac{1}{2} N_n$\;, \;
 $D_n^m = D_n^f = \frac{1}{2} D_n$\;, \;
 $C_n^m = C_n^f = \frac{1}{2} C_n$
 where ``m'' and ``f'' stand respectively for male and female for
 the $D_n$ (Dvidians) and the rest $C_n = N_0 - D_n$ .

 We continue to use the normalized population numbers $D_n$,\; $C_n$\;
 and $N_0$ eliminating $g$ from the equations since it doesn't affect the
 distribution.

\noindent
 Then the normalized population ratios are:
 \begin{align}
 \label{eq26}
 &\frac{D_n^m}{N_0^m} = \frac{D_n^f}{N_0^f} = \frac{D_n}{N_0}= \frac{\tilde D_n}{N_n} \equiv
 r_n\notag\\
 &\frac{C_n^m}{N_0^m} = \frac{C_n^f}{N_0^f} = \frac{C_n}{N_0} = \frac{\tilde
 C_n}{N_n}=
 \frac{N_0 - D_n}{N_0}= 1-r_n\,.
 \end{align}

 Now we sort the couple types and count them with their
 probabilities.  (One can envision it again by the
 two card boxes described above or by a Roulette wheel
 procedure: The $D_n^m$ and $C_n^m$ are randomly marked in the
 cycling wheel and the $D_n^f$ and $C_n^f$ are randomly marked in
 the stationary platform.)

\noindent
 The couple types and their probabilities are:
 \begin{equation}
 \label{eq27}
    \begin{array}{llll}
     (1) & D_n^m\to D_n^f & : & D_n^m r_n=\frac{N_0}{2} r_n^2 \\
     (2) & D_n^f\to C_n^f & : & D_n^m (1-r_n) = \frac{N_0}{2} r_n(1-r_n) \\
     (3)& C_n^m\to D_n^f & : & C_n^m r_n = \frac{N_0}{2} (1-r_n) r_n \\
     (4) & C_n^m\to C_n^f & : & C_n^m(1-r_n) = \frac{N_0}{2} (1-r_n)^2 \\
   \end{array}\,.
  \end{equation}

\noindent
  The first three types give for each couple two D-children, and the
  fourth one gives zero D-children, giving altogether for the next generation:
  $$
  D_{n+1} = N_0 (2-r_n) r_n = \left(2-\frac{D_n}{N_0}\right) D_n
  $$
  which is the same equation as Eq. \eqref{eq23}.

\vspace{1.3cm}

\noindent
 \textbf{Upper bound estimate:}

\vspace{0.3cm} \noindent
 We would have liked to solve
\eqref{eq23} recursively for $ED_n$ up to $n=N$. This however cannot
be done since:
\begin{equation}\label{eq28}
 E(D_n^2) = (ED_n)^2+ E (D_n - ED_n) ^2 \,.
  \end{equation}
and hence (Jensen's inequality)
  $$ ED_n^2 \ge (ED_n)^2 \,.$$
The same holds for conditional expectations. Then by Eq.
\eqref{eq23}:
 \begin{equation}\label{eq29}
    E(D_{n+1}) \le  2ED_n - \frac{E^2(D_n)}{N_0}\,.
  \end{equation}
Therefore, we can solve recursively the equation:
\begin{equation}\label{eq30}
   E\,\hat{D}_{n+1} = 2\, E\hat{D}_n - \frac{(E\,\hat{D}_n)^2}{N_0}
\end{equation}
but the solution will be an \textit{upper bound} to the solution of
\eqref{eq23} for $ED_n$.

In the next section we will show that $(ED_n)^2$ is a very good
approximation to $E(D_n^2)$ and consequently Eq. \eqref{eq24} will
be a very good approximation for $N_0>>1$.

\vspace{0.3cm}

\section{The random walk model for $E[(D_{n+1} - D_n)| D_n]$}

We wish to evaluate the standard deviation of our statistical
process at the $n$th generation that leads to the descendant
population of the $(n+1)$ generation. It is needed for evaluating
Eq. \eqref{eq28}, and for that we apply a random walk model
\cite{2,4}.

Consider the unordered collection of the pairs of cards described in
section 2, but without those cards where both couple members are
non-descendants. It will yield $D_n$ pairs. Next arrange these $D_n$
pairs randomly and independently as a sequence running from 1 to
$D_n$. Let $i, \; 1\leq i \leq D_n$ denote this sequence. Card $i$
can have on it a pair of Davidians or one Davidian and one
non-Davidian. Set $\lambda (i)=2$ in case that a Davidian married a
non=Davidian and $\lambda (i)=1$ if Davidian married a Davidian. The
sequence $\lambda (i), i=1,2,...,D_n$ constitutes a random walk
process of $D_n$ moves with a probability $(D_n/N_n)$ to make 1 step
($\times$ 1 D descendants), and a probability $(1- D_n/N_n)$ to make
2 steps ($\times$2 non-D descendants). The total average distance of
the walk gives of next generation Davidian descendants $D_{n+1}$,
given by Eq. \eqref{eq23}. A similar procedure can be attributed for
$D_{n+1} - D_n$ (successive years difference) obtained from Eq.
\eqref{eq23}:
\begin{align}\label{eq41}
    E[(D_{n+1} - D_n)|D_n)] & = \left(1-
    \frac{D_n}{N_0}\right)D_n \notag\\
    & = \left[ 1\cdot \left(1-\frac{D_n}{2N_0} \right) + (-1) \cdot
    \frac{D_n}{2N_0}\right] D_n\,.
\end{align}
Here the process is mapped to the standard random walk model where
each move is of $\pm1$ (one forward or one backward) step. The total
number of moves is $D_n$, each is either a unit step to the right
with a probability $p=(1-D_n/2 N_0)$ or a unit step to the left with
a probability $q=D_n/2N_0$.

The standard deviation of $(D_{n+1} - D_n)$ of this random walk
process is known to be given by \cite{2,4}:

\begin{align}\label{eq42}
    E^{1/2} \left\{ \Bigl[ (D_{n+1} - D_n) - E [(D_{n+1} -
    D_n)|D_n]\Bigr]^2|D_n\right\} & =
    \sqrt{D_n pq} \notag\\
    &= \sqrt{ D_n \left(1-\frac{D_n}{2N_0}\right)
    \left(\frac{D_n}{2N_0}\right)}\;\;;
\end{align}
but the left hand side of this equation is equal to:
\begin{equation}\label{eq43}
    E^{1/2} \left\{\Bigl[D_{n+1} - E(D_{n+1}|D_n)\Bigr]^2 \
    | D_n \right\}\;
\end{equation}
which is the standard deviation of $D_{n+1}$.

\noindent
Therefore:
\begin{align}\label{eq44}
    \frac{E^{1/2} \left\{\Bigl[D_{n+1} - E (D_{n+1}|D_n)\Bigr]^2
    | D_n \right\}}{E(D_{n+1} | D_n)} &= \frac{\sqrt{D_n
    \left(1-\frac{D_n}{2N_0}\right)
    \left(\frac{D_n}{2N_0}\right)}}{\left(2-\frac{D_n}{N_0}\right)
    D_n} =
    \sqrt{\frac{c}{N_0}} \;,
\end{align}
where $c\in (1/8,\,1/4]$ is a number of order 1.

\noindent
Then from Eq. \eqref{eq28} we have:
\begin{equation}\label{eq45}
    E(D_{n+1}^2 | D_n) = E^2( D_{n+1}|D_n) \,\Bigr(1+
    \frac{c}{N_0}\Bigr)\,,
\end{equation}
valid for any $n$.
 \noindent
Therefore $E^2 D_n \approx ED_n^2$ and we will have for the
recursive equation \eqref{eq23}:
\begin{equation}\label{eq46}
E(D_{n+1}) \approx \left(2-\frac{ED_n}{N_0}\right)E D_n \,.
\end{equation}
\noindent Then the descendant ratio $r_n=\frac{E\tilde D_n}{N_n} =
\frac{ED_n}{N_0}\;$ (and $r_0= \frac{1}{N_0}\;$) will be given by:
\begin{equation}
 \label{eq47}
r_{n+1} \approx (2-r_n) r_n
 \end{equation}
The last two equation are \textbf{\emph{key results}} from which we
derive our main conclusions.

It should be noted that the accumulated error in the recursive
process along $n$ (in our cases $n=1,2,...,50$), is negligible since
the relative error we found for a step in $n$ is very small, of the
order of $1/N^{1/2}_0$ (for any $n$), and $N_0 \sim 10^3-10^9$.

\vspace{0.4cm}

\section{Examples for the recursive equation solution}

We give below examples with plots of the descendant population ratio
$r_n \approx \frac{E(\tilde D_n)}{N_n} = \frac{E(D_n)}{N_0}$ along
the generation number, given by the exact solution of the recursive
equation \eqref{eq47}: $ r_{n+1} \approx(2-r_n) r_n $.

The plots in Fig.~1 give the $r_n$ dependance on $n$ for various
values of $N_0$ (corresponding to various $r_0=1/N_0)$. For a
population of $10^5$, we get for the DAD full spreading time (the
number of generations for the whole population to becomes
descendants) about 20 (or $20\times 25=500$ years). For each
additional factor of $10$ in $N_0$ we need $1/\log_{10} 2 = 3.22$
more generations. Thus even for a population of $N_0 = 10^9$ it
takes only $(9/\log_{10} 2) + 4 = 33.9$ generations (850 years) to
the DAD full spreading time.

\begin{figure}[h]  
    \centerline{\includegraphics[scale=.18]{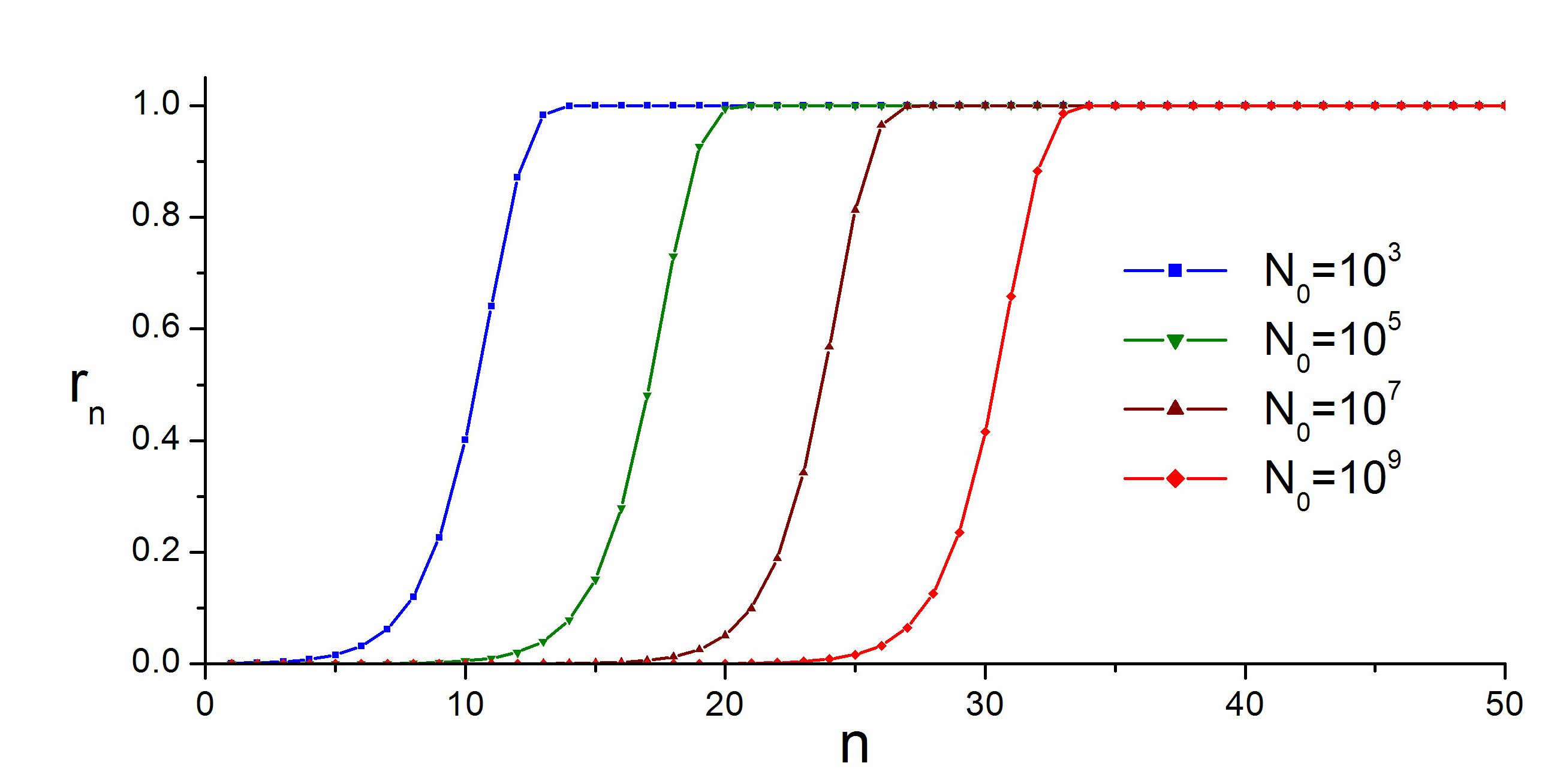}}
  \caption{Descendant population ratio $r_n$ vs. $n$, obtained from the recursive
equation - Eq. \eqref{eq47}, for various starting population
numbers: $N_0 = 1/r_0 \,=10^3, 10^5, 10^7$ and $10^9$,
(corresponding to plots from left to right, respectively).}
\end{figure}

\vspace{0.2cm}

\section{Approximating the recursive equation with a \newline differential equation}

We derive from the recursive equation \eqref{eq46} an approximated
differential equation.  It will have the advantage that is possess
an explicit solution, although the fitting is not exact.

\noindent We have from Eq. \eqref{eq46}:
\begin{equation}\label{eq51}
    ED_{n+1} - ED_n \approx \left(1 - \frac{ED_n}{N_0}\right)ED_n\,.
\end{equation}
Replacing $ED_n$ with a continuous and differentiable function
$\varphi(n)$, we obtain the equation:
\begin{equation}\label{eq52}
    \frac{d\varphi(n)}{dn} = \left(1-
    \frac{\varphi(n)}{N_0}\right)\varphi(n)\,.
\end{equation}
Hence
\begin{equation}\label{eq53}
    dn=\frac{d\varphi(n)}{\varphi(n)
    \left(1-\frac{\varphi(n)}{N_0}\right)}\,.
\end{equation}
Set $r= \frac{\varphi(n)}{N_0}, \; r\in (0,1)$ and $r_0 = 1/N_0 \;$,
then
\begin{equation}\label{eq54}
    \int_{r_0}^r \frac{dr}{r(1-r)} = \int_0^n dn\,\,.
\end{equation}
Hence
\begin{equation}\label{eq55}
    n=\ln\frac{r/r_0}{(1-r)/(1-r_0)} = \ln
    \left(\frac{N_0r\left(1-\frac{1}{N_0}\right)}{1-r}\right)\,.
\end{equation}
Solving for $r$ yields
\begin{equation}\label{eq56}
   r(n) = \frac{e^n}{e^n + N_0 -1}\;\;.
\end{equation}
This solution is only a rough estimate for the recursive equation -
Eq. \eqref{eq47}, and deviates from its exact solution especially
for low $n$ including the transition region to $r \rightarrow 1$. By
rescaling $n\to n \ln 2$, we obtain a modified equation:
\begin{equation}\label{eq57}
    r(n) = \frac{2^n}{2^n + N_0 - 1} \;\;,
\end{equation}
that describes the initial descendant evolution more accurately.
However, it increases the width of the transition region. Eqs.
\eqref{eq56} and \eqref{eq57} provide a rule of thumb approximation
that $r(n)= 1/2$ is obtained at $2^n \approx N_0$.

Fig.~2 shows plots of the solutions - Eqs. \eqref{eq56} and
\eqref{eq57} of the two differential equations which are compared to
the exact solution given by the recursive equation - Eq.
\eqref{eq47}.

\begin{figure}[h]  
    \centerline{\includegraphics[scale=.13]{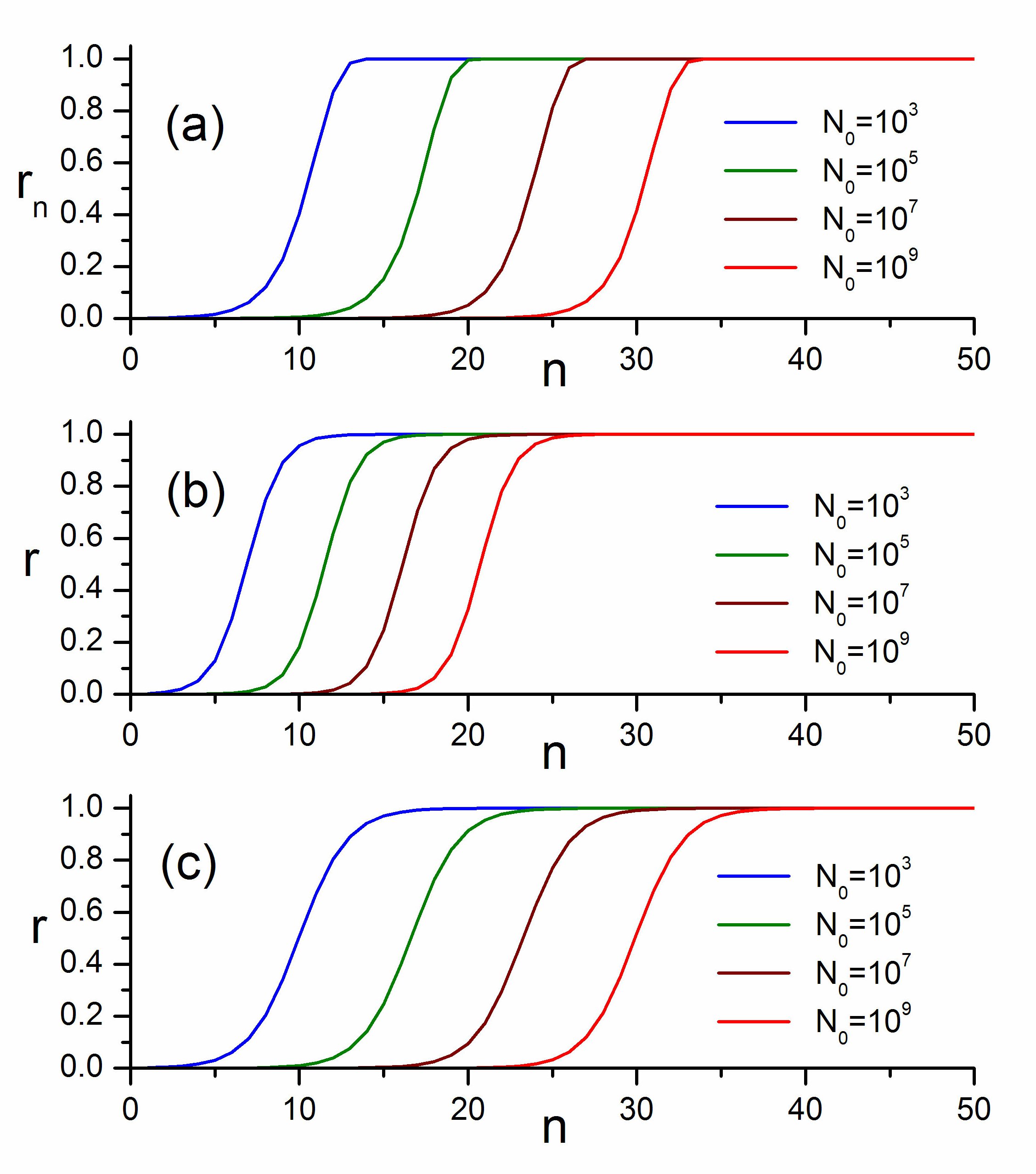}}
  \caption{Descendant population ratio $r_n$ obtained from: (a) the
 recursive equation \eqref{eq47},
  and the approximations (b) Eq. \eqref{eq56}, and (c)
  Eq. \eqref{eq57}}.
\end{figure}

\vspace{0.1cm}

\section{Dynasty Extinction}
Our model uses a growth factor that can vary from generation to
generation, but takes a fixed average number of children of
$m\,(=2g)$ per family for the whole population. The overall DAD
model that we presented gives an accurate description of the
descendant distribution, but as we said, it needs attention to the
first few generations. Then, when the dynasty starts to build up
($D_n$ is a small number) it has fragility features just as the
Galton-Watson (GW) process \cite{1,2}. The GW process was presented
to find the stability and the extinction of surnames (or families)
that is dependent on the average number of sons (or children) per
family. We already mentioned that we need $m>1\;\, (g>1/2)$, just as
in the GW process, to eliminate the possibility for the D
extinction. However even for larger $g$ there is a certain nonzero
probability that a dynasty disappears. We need for that the
probabilities for the various number of children per two parents
that includes a finite probability for having no children. The zero
children probability causes the extinction. We give here a brief
analysis that basically follows the GW process.

For a Poissonic distribution the probability for the number of
children $k$ per family is $p_k= m^k e^{-m}/k!$. For example, for
$m=2\;(g=1)$: $p_0=0.135,\;p_1=0.27,\;p_2=0.27$, and $p_3=0.18$. It
is easy to realize that most of the extinction probability comes
from a descendant chain of successive families along the generation
line $n$, each with one child, terminated by a zero children family.
The probabilit`y for that is the sum of the geometric series
$\sum_{n} p_1^n p_0$=0.185. For $m=3$ the sum is $0.058$. We will
see below that these numbers are very close to the results obtained
by the GW calculation. In the GW process for Poissonic offspring
distribution, the extinction probability can be obtained from the
probability generating function \cite{2}: \vspace{-0.2cm}
\begin{equation}\label{eq61}
\tilde{s}_{n+1} = e^{m(\tilde{s}_n-1)}\,,
\end{equation}
\noindent and the survival probability $s_n\equiv 1-\tilde{s}_n$ is
given by:\vspace{-0.2cm}
\begin{equation}\label{eq62}
s_{n+1} = 1-e^{-ms_n}.
\end{equation}
\noindent For $m \leq 1\;\;(g\leq 1/2)$ this recursion levels off to
zero (i.e. family extinction). For $m > 1$ it levels off to a
nonzero value given by the nontrival fixed point of Eq.
\eqref{eq62}: one of the two roots of $s=1- e^{-ms}$. The trivial
fixed point is always $s_a=0$, and the second one $s_b$, that
depends on $m$, gives the survival probability. [When starting the
recursion at $s_0=1/N_0$ it gives for this process the evolution
along $n$ of the descendant population ratio that asymptotically
reaches $s_b$.] For $m=2 \;(g=1)$ children per family, the survival
ratio is $s_b=0.797$ (extinction ratio of $0.203$). For
$m=3\;(g=1.5)$, the survival ratio is $0.94$ (extinction ratio of
$0.06$) and for $m=4\;(g=2)$ it is $0.99=99\%$ (extinction ratio of
$0.01=1\%$). The value of $m$ can vary along the time and from place
to place; the long term global population growth factor, however, is
not much higher than $g=1$. For $g \in (1,1.2)$, i.e. $m \in
(2,2.4)$, the global family survival ratio is $s_b \in(0.8, 0.88)$ .

The conclusion is that a certain fraction $(1-s_b)$ of the
population at the $n=0$ generation will eventually have no
descendants. The extinction that can occur in the first few
generations is related to the GW process. We note what is discussed
in the next section, that the extinction part doesn't affect the
basic DAD process that is governed by the average $g$ for those (the
most) who survive. Nevertheless, it means about being a descendant
of King David (or of anybody else) that: \emph{either all Israelites
are his descendant or there isn't even a single descendant;} or:
\emph{if there is for sure one descendant, then the whole population
are descendants}. We said that for a moderate population growth the
extinction part is $(10-20)\%$, meaning that there are no
descendants for that fraction of the ancestors population.
Nevertheless, as we said in the introduction, it is very likely that
King David's dynasty survived the first few generations and then the
DAD process ensured that they fully spread to the whole population.

\vspace{0.0cm}

\section{Forward vs. backward approaches; \\ Descendants vs. ancestors distributions}

Our model was constructed by going from an ancestor to his
descendants and thus look at the descendant distribution. Then it is
possible to follow the path down from generation to generation, but
the number of children per couple can vary and have some average
value of $2g$. The Galton-Watson (GW) process \cite{1,2} is an
example to such analysis. We note the basic difference between the
GW process and ours. The DAD process depends on the question to whom
a Davidian is married. A non-Davidian mate contributes $m\, (=2g)$
children per two parents, while an intra tribe marriage (Davidian to
Davidian) gives only half of it, $m/2=g$ children. In the GW like
process this question is irrelevant. Its focus is on the
distribution of the number of children per family that affect the
descendant statistics, but it doesn't include the interplay between
intra and inter tribe marriage. Therefore the GW process is relevant
in the DAD analysis in the first generations only to understand the
fragility and extinction. At that stage the DAD process is anyhow
similar to the GW one since the probability for a non-Davidian mate
is almost $100\%$. This is the reason why we could use in Section 6
the GW analysis for evaluating the survival ratio.

It is also possible to go along the opposite direction from a
descendant back in time to his ancestors. Then the route follows the
ancestors distribution rather than the descendants. The backward way
is more common because of the simplicity to track genealogical
family trees of more recent generations, and since then in the
analysis the number of parents is a fixed value of 2. Nevertheless
the different backward trajectories that can pass through the same
ancestors have probabilistic aspects. This viewpoint alludes on
another aspect, the number of paths that connect a descendant at the
$n-$th generation to a specific ancestor. This issue depends on the
growth factor and the degrees of inter versus intra community
mobility. It is more likely that the connection path number is
significantly higher within communities with a common
social-geographic history. Derrida et al. \cite{5,6} took the
backward direction analysis and obtained much understanding on
various sides of the process. We mention here only one of their
results on the population ratio of ancestors at a backward
generations given in Refs. \cite{5,6}. The result there is related
to the GW process giving for the non-ancestors population ratio at a
backward generation $i$ the recursive equation: $\tilde{s}_{i+1} =
e^{m(\tilde{s}_i-1)}$. For the ancestor ratio $s_i\equiv
1-\tilde{s}_i$, it becomes $ s_{i+1} = 1-e^{-ms_i}$ (dependent on
$m$ or $g$), which is equivalent to Eq. \eqref{eq62} of the GW
process, and so it differs from our approach.

\vspace{0.0cm}

\section{Conclusions}
We have presented a model for the distribution of descendants along
generations, the DAD route. The descendant population and the ratio
are given by the recursive equations \eqref{eq46} and \eqref{eq47}.
The descendant ratio, given for a few examples in figure 1, is shown
to reach 1 in a relatively few generations. For an initial
population of $N_0$ the DAD full spreading time is $\sim\log
N_0/\log2 + 4$, that gives about 20 generations (500 years) for $N_0
= 10^5$. Every additional factor of $10$ in $N_0$ adds $1/\log_{10}
2 = 3.32$ generations to the DAD full spreading time. The basic DAD
route behavior, in particular the descendant population ratio does
not depend on the population growth factor $g$, but only on the
initial population $N_0$.

We have not included here any genetic or geo-social-demographic
aspects. It is clear that DAD will not spread into and out of very
isolated groups. Nevertheless, we saw how quick the spreading
process is. For a small group of $10^3$ or $10^4$ it takes about 14
and 17 generations (350 and 425 years) to reach a full descendant
spreading ratio. For all Israelites at King's David time it took
24-26 generations (600-650 years) to make every Israelite his
descendent. That therefore happened already at 400 BC, the beginning
of the era of the Second Temple in Jerusalem. Even for the whole
world population at King's David time (1000 BC), estimated as
$5\times 10^7$, it is but 29.5 generations (740 years). Segregation
of local communities can slow down the process, but only in a
limited way for relatively short time. It would be sufficient that
one descendant migrates to another community, say with a similar
population number, to make the whole population descendants in a few
generations. Therefore more globally one might say that even if
mankind started with many Adams and Eves, it took relatively very
short time to consider the whole founder group at any early era, as
a Common Forefather -- Super DAD -- of all of us.

For the future, the DAD route means that, assuming a reasonable
population mobility, each of us on earth today ($N_0=\sim 5\times
10^9)$ (beyond the low extinction percentage discussed in Section 6,
and assuming that no catastrophic event happens, will be an ancestor
of everyone in the world - on Earth and beyond?) in $\sim \log
(5\times 10^9)/\log2+4 \approx 36$ generations ($\sim$ 900 years)
from now. Can this picture lend a philosophical meaning to what is
said about a common forefather of mankind?  We saw that all of us
have common ancestors and eventually we will be ourselves the
ancestors of everyone in the future in a relatively short period of
time. They all were our Fathers and Mothers and they all will be our
Sons Daughters\ldots

We finally note that after we first deposited the paper in ArXiv,
our attention was drawn to the work of Derrida et al. on the same
subject \cite{5,6}. We believe that our paper adds new insight and
results treating the process along the generation line rather than
the backward way. We have added a short discussion on the difference
between the approaches in Sec. 7.

\vspace{0.0cm}

\noindent \textbf{Acknowledgement:} We wish to thank Israel
Bar-David, Nahum Shimkin, Adam Shwartz, and Ofer Zeitouni for useful
remarks. We also thank Yosi Avron and Yosef Maruvka for drawing our
attention to the work of Bernard Derrida et al. on the subject.

\end{document}